# HYBRID FUZZY LOGIC AND PID CONTROLLER FOR pH NEUTRALIZATION PILOT PLANT

Oumair Naseer[1], Atif Ali Khan[2]

[1,2] School of Engineering, University of Warwick, Coventry, UK,
o.naseer@warwick.ac.uk
atif.khan@warwick.ac.uk

## ABSTRACT

*Use of Control theory within process control industries has changed rapidly due to the increase complexity of instrumentation, real time requirements, minimization of operating costs and highly nonlinear characteristics of chemical process. Previously developed process control technologies which are mostly based on a single controller are not efficient in terms of signal transmission delays, processing power for computational needs and signal to noise ratio. Hybrid controller with efficient system modelling is essential to cope with the current challenges of process control in terms of control performance. This paper presents an optimized mathematical modelling and advance hybrid controller (Fuzzy Logic and PID) design along with practical implementation and validation of pH neutralization pilot plant. This procedure is particularly important for control design and automation of Physico-chemical systems for process control industry.*

## 1. INTRODUCTION

Control system design for $pH$ neutralization pilot plant has a long history, due to its non-linear characteristics, uncertainty and large number of requirements from environment legislation which revised constantly. Most of the classical control techniques are developed on the bases of linear theory. When these techniques are applied to the chemical systems having kinetic reactions and thermodynamic relationships, they do not provide the adequate system performance and hence fails to capture the entire operating range. $pH$ neutralization process mainly consists of $pH$ measurement of an acid-base chemical reaction in which hydrogen ions and hydroxide ions are neutralized or combined with each other to form water, while the other ions involved remained unchanged. Acid is a substance that ionises in water to produce hydrogen ions and base is a substance which ionises in water to produce hydroxyl ions. The characteristics of acid-base neutralization reaction are normally represented by a titration curve. Fig.1 below provides the information about the equilibrium point [1, 2], the type of acid-based (strong/weak) involved and the total volume or amount of substances involved at the end of the titration process. The S-shaped curve shown in Fig. 1 depends on the concentration and composition of acid and based used in the process reaction. It can be seen that at $pH$ value of 7, a small change in input produces a large change in output, which provides the significance of controlling $pH$ Value [3, 33, and 35].

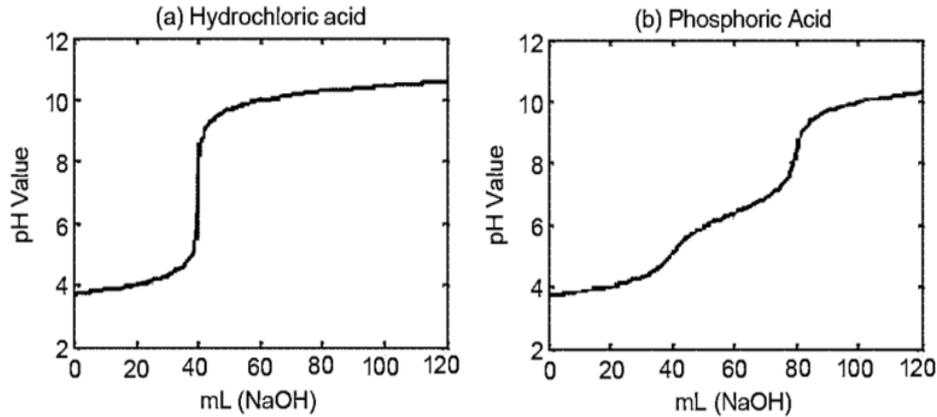
Figure 1: Titration curve for acid-base process reaction.

Scale of measuring acidity of the system ranges between $pH$ values 1 to 14. At room temperature $25°C$, if the $pH$ value is less than 7, the mixed solution has higher concentration of hydrogen ions; hence it is acidic in nature. If $pH$ value is greater than 7 then the mixed solution has higher concentration of hydroxyl ions and is alkaline/base in nature. However, if $pH$ value is 7 then the mixed solution is neutral. For industrial safety, all waste water should maintain a $pH$ level of 7±1. $pH$ control loop mainly consists of (i) an open loop type of control scheme; in which the control valve opening is kept at certain positions for specific time durations, (ii) a feedback control scheme; which involves a direct relationship between the control valve opening and the $pH$ value in the process, and (iii) a feed-forward control scheme; in which controller will compensate for any measured disturbance before it affects the process (i.e. the $pH$ value in the case of this application). In the past few years, several control strategies have been developed in the process control to improve the performance of the system. This is achieved by efficiently defining an optimal $pH$ pilot plant model within the control structure and this trend keep on increasing day by day. Classical control schemes such as Proportional-Integral-Derivative (PID) control, depends on the optimal tuning of control parameters which are proportional gain, integral gain and derivate gain. On the other hand, performance of fuzzy logic control depends on the vigilant selection of the membership function for the input set and output set parameters. Traditional PID or Fuzzy logic based controller cannot provide optimal performance as compared to the combination of PID and Fuzzy logic or the combination of adaptive neural network control and fuzzy logic control. In control system architecture, integration of such hybrid control provides new trend towards the realization of intelligent systems. In this paper an optimal mathematical modelling of $pH$ neutralization pilot plant with Hybrid controller (Fuzzy logic and PID) design, implementation and validation is presented.

## 2. RELATED WORK

A rigorous and generally applicable method of deriving dynamic equations for $pH$ neutralization in Continuous Stirred Tank Reactors (CSTRs) is presented in [5, 6]. In [4], author uses the same model and provides a more optimal solution for adaptive control for $pH$ neutralization process control. A new concept concerning the averaging $pH$ value of a mixture of solutions is introduced in [7, 8]. It gives the idea to utilize reaction invariant variables in calculating the $pH$ value of mixtures of solutions, instead of using a direct calculation involving a simple averaging of hydrogen ions. In [9], authors introduced a systematic method for the modelling of dynamics of the $pH$ neutralization process. They used a hypothetical species estimation to obtain the inverse titration curve so that overall linearization of the control loop can be utilized. Fundamental properties of continuous $pH$ control are investigated in [10] using

Proportional-Integral-Derivative (PID) controller. Nonlinear adaptive control for $pH$ neutralization pilot plant is presented in [11]. In [12, 13] author evaluated the performance of the system on a bench scale $pH$ neutralization system in order to gain additional insight in terms of the practical application. A feedback based nonlinear controller is developed in [14, 30] by applying an input-output linearization approach to a reaction invariant model of the process by using a Proportional Integral (PI) controller which utilizes an open-loop nonlinear state observer and a recursive least squares parameter estimator. A mathematical modelling of the $pH$ neutralization process in a continuous stirred tank reactor which is based on a physico-chemical approach is presented in [15]. In [16], an adaptive $pH$ control for a chemical waste water treatment plant is presented. The same approach is extended in [17]. In [18], author presented a new approach for an adaptive combined feedback-feedforward control method for $pH$ control which was based on a quantitative physico-chemical analysis of the $pH$ neutralization process. However, this work has relatively high signal to noise ratio. Similar approach is further extended by using an adaptive $pH$ control algorithm in [19]. A comparison of linear and non-linear control for $pH$ neutralization plant is presented in [20]. In [21, 22], author presents the investigation of the controller performance, such as tracking of the lime flow-rate set point, investigation of different conditions of normal process operation and operation without ignition. A new approach to $pH$ control which utilizes an identification reactor to incorporate the nonlinearities of the $pH$ neutralization process is described in [23]. Indirect adaptive nonlinear control of $pH$ neutralization plant is presented in [24]. In [25, 30-32] practical control design issues for a $pH$ neutralization is investigated with the objective to design an online identification method based on use of an extended Kalman filter. However, this work has low processing power for high computational systems. In [26, 34] author outlines the framework of the controller and develops a fuzzy relational model which is based on a fuzzy logic approach. Similar approach is extended in [27]. In [28, 29], author outlines a technique in which the genetic algorithm approach is employed to alter membership functions in response to changes in the process. In [36] a fuzzy fed PID control of non-linear system is presented but this work doesn't capture the system's characteristics in terms of signal transmission delays.

## 3. $pH$ NEUTRALIZATION PILOT PLANT ARCHITECTURE

The architecture diagram of $pH$ neutralization pilot plant is shown in Fig 2. This plant is assembled by using the state of the art industrial instruments and measuring systems.

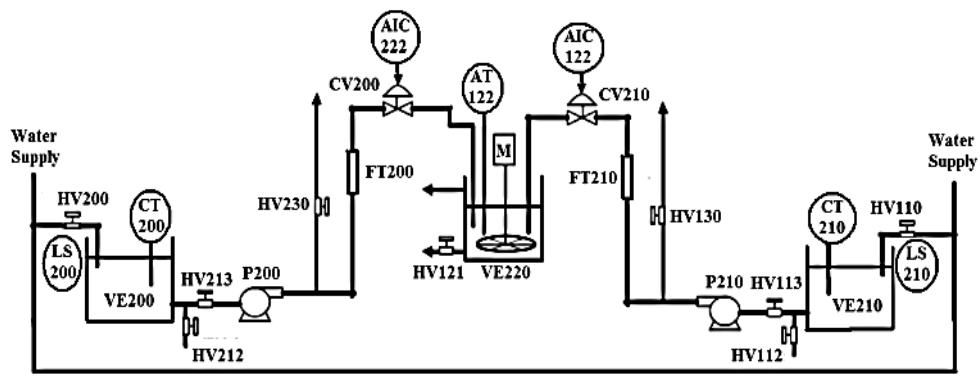

Figure 2: $pH$ Neutralization pilot plant architecture.

This pilot plant consists of three tanks; VE210 is the alkaline tank, VE200 is the acidic tank and VE220 is the mixer tank. P210 and P200 are the pumps which are used to get the desired amount of acid and base in the mixer tank. FT 210 and FT 200 are the flow transmitters. The job of the flow transmitters is to transmit the precisely measured amount of acid and base respectively to the central computing system. Two control valves CV210 and CV200 are connected to the alkaline and acid pipelines respectively to control the flow of the acid and base getting into the mixer tank. These control valves are fully digital and are operated by using electrical signals. The mixer tank contains a continuous steering motor, which keeps the solution in the mixer tank at the uniform state. Because whenever a corresponding amount of acid or base is added into the mixer tank, a certain amount of time is required in order for the chemical reaction to take place. Continuous steering ensures the stability of $pH$ level at each point in the mixer tank. The overall system architecture diagram is shown in Fig. 3.

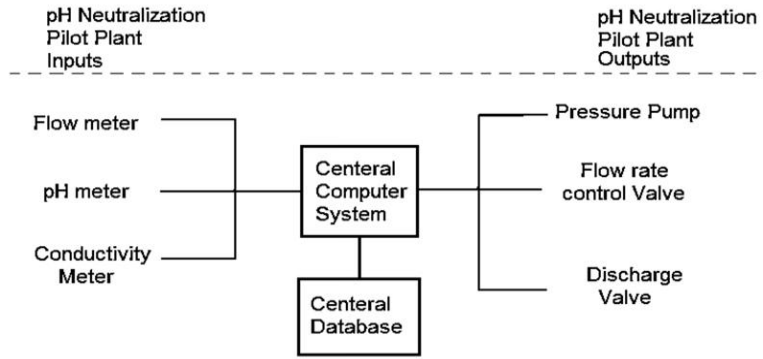

Figure 3: Overall system architecture.

Inputs to the system are the data coming from the $pH$ neutralization pilot plant i.e. the flow meter sensors, conductivity monitor sensors, $pH$ value meter. The corresponding outputs are the actuators that control the flow valves. Processor senses the $pH$ level of the mixed solution in mixer tank and controls the valves of the acid and base tank accordingly to maintain the $pH$ level equals to 7.

## 4. MATHEMATICAL MODELLING OF $pH$ NEUTRALIZATION PILOT PLANT

Acid used in mixer tank is $H_2SO_4$ and alkaline used is $NaOH$. Table.1 shows the various process variables used in the neutralization plant.

Table 1: Process variables for $pH$ neutralization pilot plant.

| No: | Process Variables | Instruments |
|---|---|---|
| 1 | $pH$ value form mixer tank | $pH$ Meter |
| 2 | Concentration of alkaline tank | Conductivity Meter |
| 3 | Concentration of acid tank | Conductivity Meter |
| 4 | Flow rate of alkaline stream | Flow meter |
| 5 | Flow rate of acid stream | Flow meter |

Mixer Tank is the continuous steering tank reactor. F1 is the flow rate of the acid and F2 is the flow rate of the alkaline. C1 is the concentration of acid and C2 is the concentration of alkaline.

The mathematical equation for mixer tank can be defined as the rate of accumulation of non-reactant species (within element volume) is equal to the rate of flow of non-reactant species (into element volume) minus the rate of flow of non-reactant species (out of element volume). This can be written as:

$$V\frac{d\alpha}{dt} = F_1 C_1 - (F_1 + F_2)\alpha \quad (1)$$
$$V\frac{d\beta}{dt} = F_2 C_2 - (F_1 + F_2)\beta \quad (2)$$

Where, V is the volume of the tank. $\alpha$ and $\beta$ are the non-reactant components of the system for acid and alkaline respectively. They are defined as:

$$\alpha = [H_2SO_4] + [HSO_4^-] + [SO_4^{-2}] \quad (3)$$
$$\beta = [N_a^+] \quad (4)$$

Based on the electro-neutrality condition, sum of all the positive charges is equal to the sum of all negative charges and can be written as:

$$[N_a^+] + [H^+] = [OH^-] + [HSO_4^-] + 2[SO_4^{-2}] \quad (5)$$

The equilibrium constant expression for water and $H_2SO_4$ are as follows:

$$K_1 = \frac{[H^+][HSO_4^-]}{H_2SO_4} \quad (6)$$
$$K_2 = \frac{[H^+][SO_4^{-2}]}{HSO_4^-} \quad (7)$$
$$K_W = [H^+][OH^-] \quad (8)$$

$K_W$ is the constant ionic product of water and is equal to $1.0 \times 10^{14}$. $K_1$ and $K_2$ are the two acid dissociation constants for sulphuric acid with $K_1 = 1.0 \times 10^3$ and $K_2 = 1.2 \times 10^{-2}$. $pH$ value of the solution can be calculated by using the following equation:

$$PH = -\log_{10}[H^+] \quad (9)$$

Equation 5 can be solved for the value of Hydrogen ions H+ by using (6, 7, 8, and 9) and can be written as:

$$[H^+]^4 + a_1[H^+]^3 + a_2[H^+]^2 + a_3[H^+]^1 + a_4 \quad (10)$$
$$a_1 = K_1 + \beta$$
$$a_2 = \beta K_1 + K_1 K_2 - K_W - K_1 \alpha$$
$$a_3 = \beta K_1 K_2 - K_1 K_W - 2K_1 K_2 \alpha$$
$$a_4 = -K_1 K_2 K_W$$

Equation 10 is the Physico-chemical $pH$ neutralization equation for the mixer tank. The block diagram of the resulting plant is shown in the Fig. 4.

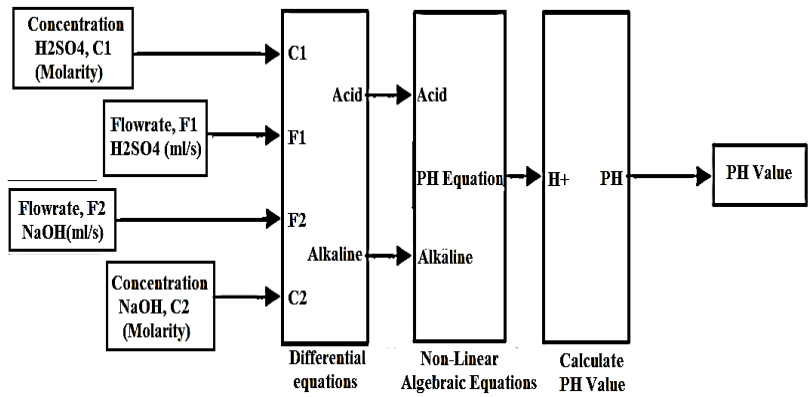
Figure 4: Mathematical model of $pH$ neutralization pilot plant.

Fig. 5 shows the dynamic response of the system which is clearly a non-linear system. $pH$ value starts from 3. The variation of $pH$ value from 3-4 is low and from 4-6 variation is very high. However, from $pH$ values 6-8 the slope is quite linear and the variation is moderate. From $pH$ values 8-10, variation is again higher and finally from $pH$ values 10-12 variation is low.

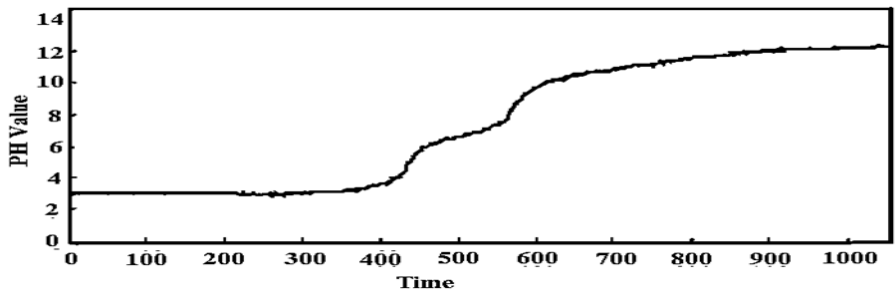
Figure 5: Dynamic response of $pH$ neutralization pilot plant.

## 5. PROPORTIONAL INTEGRAL DERIVATIVE (PID) FLOW-RATE CONTROL DESIGN

The characteristics curves for flow rates of acid and alkaline valves are shown in Fig. 6. It is evident that flow control for up scaling (opening of valve) and down scaling (closing of valve) is not the same. The error varies from 2% to 6%. PID controller is designed and tuned to capture these variations [3].

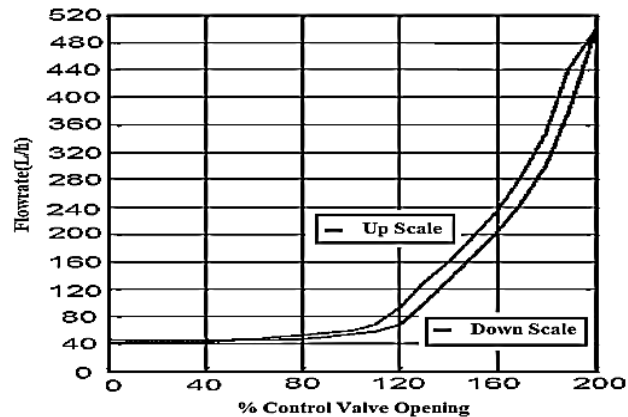
Figure 6: Flow-rates of acid and base streams.

Firstly, the proportional gain is set to a minimum value and the other parameters integral and derivative terms are set to give zero action. The proportional gain is then gradually increased until oscillations start to appear in the measured closed-loop system response. The gain is then adjusted so that the oscillations maintain constant amplitude. The value of gain that is used to achieve this condition is termed as ultimate proportional gain with value (G=18) at the period (P=33). Based on Table 2 proportional gain Kp = 10.8, integral gain Ki = 0.65 and derivative gain Kd = 44.5.

Table 2: Ziegler-Nichlos tuning formula for a closed loop system.

| Type of Controller | P | PI | PID |
|---|---|---|---|
| Proportional Kp | 0.5 G | 0.45G | 0.6G |
| Integral Ki | - | 1.2Kp/P | 2Kp/P |
| Differential Kd | - | - | KpP/8 |

## 6. FUZZY LOGIC CONTROLLER DESIGN

Fuzzy logic controller mainly consists of three parts: (i) Fuzzification: process of converting system inputs (process variables) into grades of membership for linguistic of fuzzy sets, (ii) Fuzzy interference: mapping input space to output space using membership functions, logic operations and if-then rule statements and (iii) Defuzzification: process of producing quantifiable results (control valve inputs for PID controller) in the light of given fuzzy sets and membership degree. Overall performance of the fuzzy logic controller depends on the selection of the membership function of input and output sets.

The set point of the desired $pH$ value is entered manually while other process control variables are controlled automatically based on the information (feedback) coming from the plant output. The job of the fuzzy logic controller is to maintain the corresponding $pH$ value while manipulating the process control variables. When the current $pH$ value is less than the desired value, Fuzzy logic controller sets a new point for the PID valve flow rate controller. The new value of current set point depends upon the difference between the current $pH$ value of the plant reactor and the desired $pH$ value. The overall diagram of the system is shown in the Fig. 7.

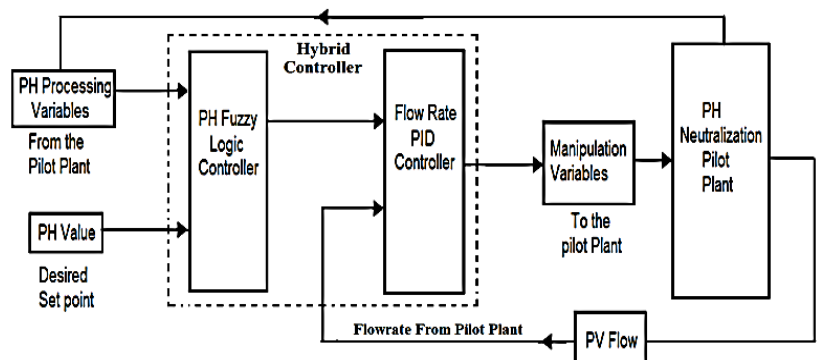

Figure 7: Logical diagram of fuzzy logic and PID controller.

Table 3 shows the membership function description and the parameters for fuzzy logic input set. The mid condition is positioned between -1 and 1 to ensure the smoothness of the desired $pH$ value and to make certain that the zero offset for the steady state is achievable. The overall performance of the system is determined by the input and output sets membership functions.

Table 3: Membership function description and parameters for input set.

| Symbols | Descriptions | Type | Parameters | | | |
|---|---|---|---|---|---|---|
| NXL | Negative Extra Large | Trapezoid | -5.0 | -5.0 | -4.0 | -2.0 |
| NL | Negative Large | Triangle | -3.0 | -2.0 | -1.0 | |
| NM | Negative Medium | Triangle | -2.0 | -1.25 | -0.5 | |
| NS | Negative Small | Triangle | -1.0 | -0.5 | 0 | |
| Z | Zero | Triangle | -0.5 | 0 | 0.5 | |
| PS | Positive Small | Triangle | 0 | 0.5 | 1.0 | |
| PM | Positive Medium | Triangle | 0.5 | 1.25 | 2.0 | |
| PL | Positive Large | Triangle | 1.0 | 2.0 | 3.0 | |
| PXL | Positive Extra Large | Trapezoid | 2.0 | 4.0 | 5.0 | 5.0 |

The entire range is divided into nine levels with the value ranging from -5.0 to 5.0. Fig. 8 further demonstrates the functionality of the input sets of fuzzy logic controller.

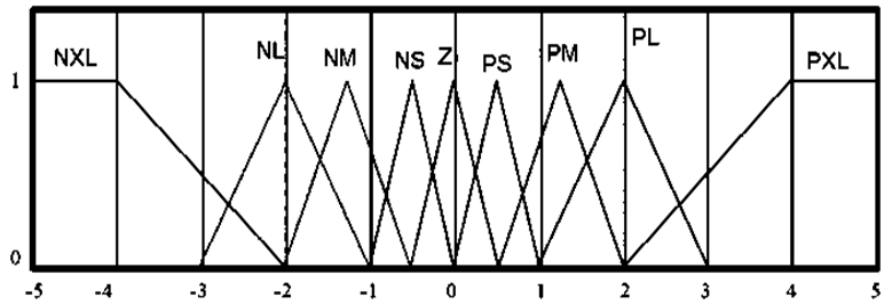

Figure 8: Demonstration of membership function of input set.

Fig. 9 shows the membership functions for the output set of Fuzzy logic controller. Output set is also divided into nine levels. The output set in this case determines the output for the PID controller (whether to increase the $pH$ value by increasing the flow-rate of acid or to decrease the $pH$ value by decreasing the flow-rate of base and vice versa) and to provide a reasonable time response.

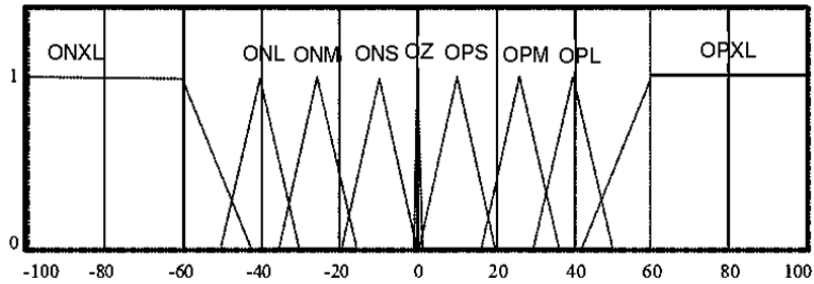
Figure 9: Demonstration of membership function of output set.

Table 4, shows the membership function description and parameters for output set. The middle range is set from -20 to 20, so that the variation in mid-condition remains minute.

Table 4: Membership function description and parameters for output set.

| Symbols | Descriptions | Type | Parameters | | | |
|---|---|---|---|---|---|---|
| ONXL | Negative Extra Large | Trapezoid | -100 | -100 | -60 | -45 |
| ONL | Negative Large | Triangle | -50 | -40 | -30 | |
| ONM | Negative Medium | Triangle | -35 | -25 | -15 | |
| ONS | Negative Small | Triangle | -20 | -10 | 0 | |
| OZ | Zero | Triangle | -0.5 | 0 | 0.5 | |
| OPS | Positive Small | Triangle | 0 | 10 | 20 | |
| OPM | Positive Medium | Triangle | 15 | 25 | 35 | |
| OPL | Positive Large | Triangle | 30 | 40 | 50 | |
| OPXL | Positive Extra Large | Trapezoid | 45 | 60 | 100 | 100 |

Table 5 defines the relationship between the input set and output set parameters of the fuzzy logic controller. This is one-to-one function for every input set parameter there is a corresponding output set variable.

Table 5: If-then rules statement description of membership function and parameters for input and output sets.

| No: | Statement | Error in $pH$ Value | Statement | Manipulated variables for PID controller |
|---|---|---|---|---|
| 1 | IF | NXL | THEN | ONXL |
| 2 | IF | NL | THEN | ONL |
| 3 | IF | NM | THEN | ONM |
| 4 | IF | NS | THEN | ONS |
| 5 | IF | Z | THEN | OZ |
| 6 | IF | PS | THEN | OPS |
| 7 | IF | PM | THEN | OPM |
| 8 | IF | PL | THEN | OPL |
| 9 | IF | PXL | THEN | OPXL |

## 7. EXPERIMENTS

First experiment is carried out to investigate the overall performance of the hybrid fuzzy logic and PID controller with the introduction of a static set point ($pH$ value = 7). For this experiment, 0.052M of H2SO4 is mixed with 0.052M of NaOH. These are the typical values for this kind of system. Two step changes are made, first at the $pH$ value of 7 and second at the $pH$ value of 10. This experiment is useful in determining the rise time (how long will it take for the output to follow the desired input) and overall response time of the system. Second experiment is carried out to investigate the robustness (to determine whether the output of the system is able to track the input or not) of hybrid Fuzzy logic and PID controller. In this experiment, different set points are introduced at different time steps in the form of a square wave. The amplitude of the wave is set to 1.5 and period is configured to 600 sec. The random range of values changes from $pH$ = 6 to $pH$ = 10. The initial $pH$ vale is set to 7. The concentration values for acid and base are set to 0.051M and 0.0489M respectively. Third experiment is performed to compare the performance of Hybrid Fuzzy Logic and PID controller against Fuzzy Logic controller. For this experiment, $pH$ value is varied from 6 to 10 at the regular intervals and the performance of both controllers is observed.

## 8. RESULTS

Fig. 10, show the results of the first experiment. Initially $pH$ value is set to 7. At time step 300 sec, the input of the system changes to $pH$ = 10. Output follows the input and at time step 380 sec, output reaches to $pH$ = 10. However, the output is delayed because the flow rate of PID controller depends on the mechanical flow valves of acid and base streams. At time step 600 sec, the input drops to $pH$ = 7 but output $pH$ value starts to drop down at 610 sec and settles down to $pH$ = 7 at 660 sec. Rise and fall time delays of the system are different because the control valve have different rise time and fall time.

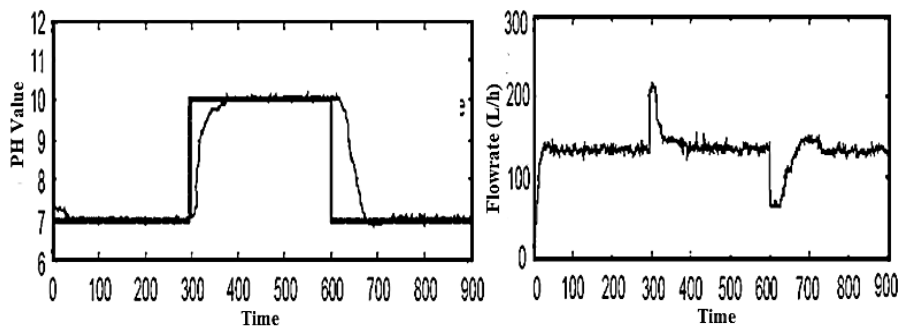

Figure 10: Experiment-1, performance of hybrid controller.

Fig. 11 shows the result of the second experiment. Output of the system follows the input square wave which shows the robustness of the controller. When input $pH$ value varies from 7 to 10, output follows the input but with the minute delay. This delayed is because the control valves require certain amount of time to maintain the desired flow rate.

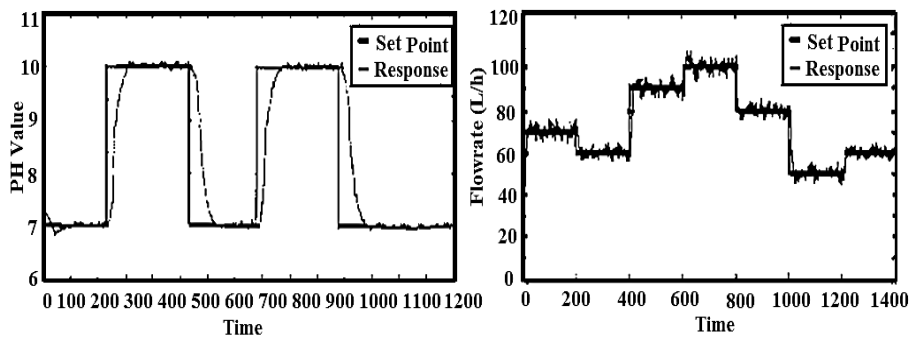

Figure 11: Experiment-2 robustness of hybrid controller.

Fig. 12, Shows the result of third experiment. It can be seen that from time intervals 1250-2000 seconds, hybrid controller is more efficient in tracking the input (set points) than Fuzzy Logic controller. Also from time interval 500-1250 seconds, Hybrid controller is more stable than fuzzy logic controller.

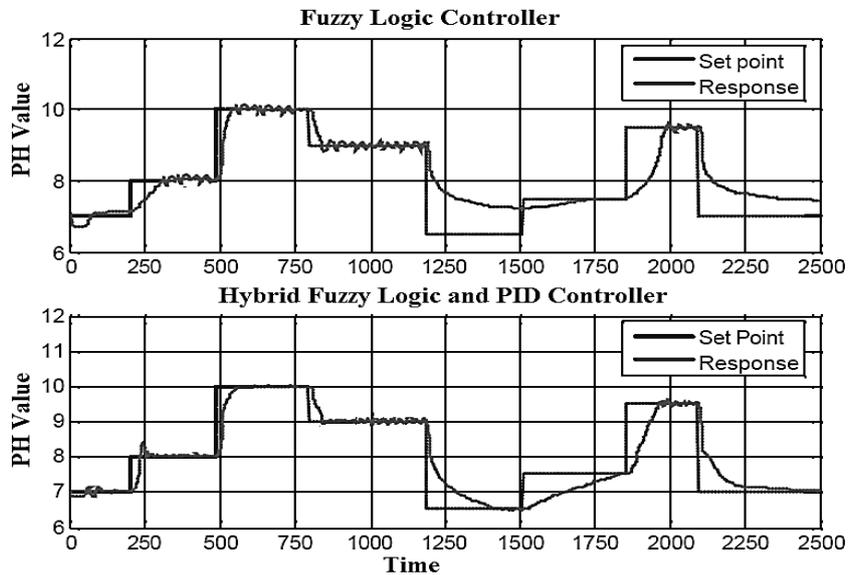

Figure 12: Experiment-3, comparison of Fuzzy Logic controller against Hybrid Fuzzy Logic and PID controller.

## 9. Conclusion and Future Considerations

This paper presents a hybrid control (PID and fuzzy logic controller) for $pH$ neutralization pilot plant. It covers the entire operating range and is more robust against the uncertainty $pH$ value variation). It is noticed that proposed hybrid controller is more stable as compared to the Fuzzy Logic controller. Process modelling approach adopted in this paper is based on the Physico-chemical principles and fundamental laws. A conventional mathematical modelling process is incorporated. Practical tests are carried out on actual system to estimate manipulating variables which were not known before the experiments. The design methodology (deriving dynamic non-linear equation) presented in this paper is generally applicable to a $pH$ neutralization plant, based on continuous steering tank reactor. The robustness of the Hybrid controller depends upon the process variables i.e. Acid/Base Flow rate control valve, $pH$ value meter, flow transmitter and concentration monitoring sensors as these instruments appear as manipulating variables during controller design and implementation.

In this paper, PID controller is used to control the flow rate of both acid and alkaline streams and a Fuzzy logic controller is used to control the $pH$ value. However in future, an adaptive neural network based controller can be used to achieve the same objective. Overall response of the system in terms of rise and fall time heavily depends on the instruments used to assemble $pH$ neutralization plant. So a more predictable $pH$ process model can be designed by using more accurate instruments.

## 10. REFERENCES

[1] Butler, J. N., Ionic Equilibrium : A Mathematical Approach Addison-Wesley Publishing, Inc., London.1964.

[2] Christian, G. D. , "Acid-Base Equilibria," in Analytical Chemistry, Sixth edn, John Wiley & Sons, Inc., United States of America, pp. 214-260. 2004.

[3] R. Ibrahim, "Practical modeling and control implementation studies on a ph neutalization plot plant", Department of Electronics and Electrical Engineering,mFaculty of Engineering,, University of Glasgow, 2008.

[4] Alvarez, H., Londono, C., de Sciascio, F., & Carelli, R., "PH neutralization process as a benchmark for testing nonlinear controllers", Industrial & Engineering Chemistry Research, vol. 40, no. 11, pp. 2467-2473, 2001.

[5] Bar-Eli, K. & Noyes, R. M., "A model for imperfect mixing in a CSTR", The Journal of Chemical Physics, vol. 85, no. 6, pp. 3251-3257, 1986.

[6] McAvoy, T. J., Hsu, E., & Lowenthals, S., "Dynamics of pH in controlled stirred tank reactor", Ind Eng Chem Process Des Develop, vol. 11, no. 1, pp. 68-78, 1972.

[7] Bates, R. G., Determination of pH: Theory and Practice, 2 edn, John Wiley & Sons, Inc, New York, 1973.

[8] Gustafsson, T. K., "Calculation of the pH value of a mixture solutions—an illustration of the use of chemical reaction invariants", Chemical Engineering Science, vol. 37, no. 9, pp. 1419-1421, 1982.

[9] Gustafsson, T. K. & Waller, K. V., "Dynamic modeling and reaction invariant control of pH", Chemical Engineering Science, vol. 38, no. 3, pp. 389-398., 1983.

[10] Jutila, P., "An application of adaptive pH-control algorithms based on physicochemical in a chemical waste-water treatment plant", International Journal of Control, vol. 38, no. 3, pp. 639-655.1983.

[11] Gustafsson, T. K. & Waller, K. V., "Nonlinear and adaptive control of pH", Industrial & Engineering Chemistry Research, vol. 31, no. 12, pp. 2681-2693, 1992.

[12] Henson, M. A. & Seborg, D. E., "Adaptive nonlinear control of a pH neutralization process", Control Systems Technology, IEEE Transactions on, vol. 2, no. 3, pp. 169-182, 1994.

[13] Bohn, C. & Atherton, D. P., "SIMULINK package for comparative studies of PID anti- windup strategies", Proceedings of the IEEE/IFAC Joint Symposium on Computer-Aided pp. 447-452, 1994.

[14] Gustafsson, T. K. & Waller, K. V., "Dynamic modelling and reaction invariant control of pH", Chemical Engineering Science, vol. 38, no. 3, pp. 389-398, 1983.

[15] Gustafsson, T. K., Skrifvars, B. O., Sandstroem, K. V., & Waller, K. V. "Modeling of pH for Control", Industrial & Engineering Chemistry Research, vol. 34, no. 3, pp. 820-827, 1995.

[16] Jutila, P. & Orava, J. P., "Control and Estimation Algorithms for Physico- Chemical Models of pH-Processes in Stirred Tank Reactors", International Journal of Systems Science,vol.12, no.7, pp.855-875, 1981.

[17]     Jutila, P., "An application of adaptive pH-control algorithms based on physicochemical in a chemical waste-water treatment plant", International Journal of Control, vol. 38, no. 3, pp. 639-655, 1983.

[18]     Jutila, P. & Visala, A., "Pilot plant testing of an adaptive pH-control algorithm based on physico-chemical modelling", Mathematics and Computers in Simulation, vol. 26, no. 6, pp. 523-533, 1984.

[19]     Gustafsson, T. K. & Waller, K. V., "Nonlinear and adaptive control of pH", Industrial & Engineering Chemistry Research, vol. 31, no. 12, pp. 2681-2693, 1992.

[20]     Bohn, C. & Atherton, D. P., "Analysis package comparing PID anti-windup strategies", IEEE Control Systems Magazine, vol. 15, no. 2, pp. 34-40, 1995.

[21]     Wright, R. A., Smith, B. E., & Kravaris, C., "On-Line identification and nonlinear control of pH processes", Industrial and Engineering Chemistry Research, vol. 37, no. 6, pp. 2446-2461, 1998.

[22]     Wright, R. A. & Kravaris, C., "On- line identification and nonlinear control of an industrial pH process", Journal of Process Control, vol. 11, no. 4, pp. 361-374, 2001.

[23]     Sung, S. W., Lee, I. B., & Yang, D. R., "pH control using an identification reactor", Industrial and Engineering Chemistry Research, vol. 34, no. 7, pp. 2418- 2426, 1995.

[24]     Yoon, S. S., Yoon, T. W., Yang, D. R., & Kang, T. S., "Indirect adaptive nonlinear control of a pH process", Computers and Chemical Engineering, vol. 26, no. 9, pp. 1223-1230, 2002.

[25]     Yoon, S. S., Yoon, T. W., Yang, D. R., & Kang, T. S., "Indirect adaptive nonlinear control of a pH process", Computers and Chemical Engineering, vol. 26, no. 9, pp. 1223-1230, 2002.

[26]     George, J. K. & Yuan B, Fuzzy Sets and Fuzzy Logic : Theory and Applications Prentice Hall, PTR, New Jersey.1995.

[27]     Kelkar, B. & Postlethwaite, B. 1994, "Fuzzy- model based pH control", IEEE International Conference on Fuzzy Systems, vol. 1, pp. 661-666.1994.

[28]     Karr, C. L. & Gentry, E. J, "Fuzzy control of pH using genetic algorithms", IEEE Transactions on Fuzzy Systems, vol. 1, no. 1, pp. 46-53.1993.

[29]     Karr, C. L., "Design of a cart-pole balancing fuzzy logic controller using a genetic algorithm", Proceeding of The International Society for Optical Engineering, vol. 1468, pp. 26-36.1991.

[30]     Gong, M. & Murray-Smith, D. J., "A practical exercise in simulation model validation", Mathematical and Computer Modelling of Dynamical Systems, vol. 4, no. 1, pp. 100-117.1998.

[31]     Murray-Smith, D. J, "Issues of Model Accuracy and External Validation for Control System Design", Acta Polytechnica, vol. 40, no. 3.2000.

[32]     Murray-Smith, D. J., "Simulation Model Quality Issues In Engineering: A Review", Proceedings 5th Symposium on Mathematical Modelling, MATHMOD Vienna, Austria, 2006.

[33]     Harvey, D., Morden Analytical Chemistry The McGraw-Hill Companies, Inc., Singapore.2000.

[34]     Jamshidi, M., Ross, T. J., & Vadiee, N., Fuzzy Logic and Control: Software and Hardware Applications Prentice Hall, Inc., New Jersey.1993.

[35]     S. Vaishnav, Z. Khan. Design and Performance of PID and Fuzzy Logic Controller with Smaller Rule Set for Higher Order System. International Conference on Modeling, Simulation and Control, San Francisco, USA. Pages 24-26: 855–858, 2007.

[36]     B. Hamed and A. El Khateb, "Hybrid Takagi-Sugeno fuzzy FED PID control of nonlinear systems," Intelligent Systems and Automation, vol. 1019, pp. 99-102, 2008.

.